\documentclass[allclo]{FBSart}
\usepackage{amsfonts}
\usepackage{amssymb}
\usepackage{amsmath}
\usepackage[dvips]{pstcol}
\usepackage[dvips]{graphicx}
\usepackage{bm}

\title{A Personal Journey Through Hadronic Exotica}
\author{Kamal K. Seth\comma\thanks{\textit{E-mail address:} 
kseth@northwestern.edu}}
\institute{Department of Physics and Astronomy, Northwestern University, Evanston, IL 60208, USA}

\runningauthor{K.K.~Seth}
\runningtitle{A Personal Journey Through Hadronic Exotica}
\begin{document}

\maketitle
\begin{abstract}
The search for exotic hadrons has been forever fascinating and challenging.  A review of many such searches, successful and unsuccessful, in which the author has been involved, is presented.
\end{abstract}

\section{Introduction}

\textbf{Exotica} and \textbf{Erotica} differ only in one letter. They are equally addictive. Like all addictions, they have consequences. They consume a lot of the resources. They make you often do things you should not do. \textbf{BUT}, they are exciting, and they give you a great surge of adrenaline.

I have to confess that over the years I have fallen for exotica, and often.  So, let me take you on a personal journey through exotica.

So, what is Exotic? Exotic has to be unexpected.  Exotic has to have the nature of the ``forbidden fruit''. Exotic in hadronic physics often begins with provocative suggestions by theorists, which drives experimentalists to search for it, often at \textbf{exotic cost} (think Higgs). At other times, it begins with an unexpected experimental observation for which theorists come up with exotic explanations (think $J/\psi$).  I want to tell the story of the hadronic exotica, necessarily from a personal point of view.

\section{Chasing Exotica in Nuclear Physics}

I began my career as a nuclear physicist.  So, my first run in with exotica was in the search for \textbf{exotic nuclei}.  Nuclei are exotic if they are very rich in neutrons, i.e., have an exceptionally large value of $(N-Z)/A$., or if they are just very heavy, $A\gg240$. In the 1970's, there were no easy ways of making a nucleus which was very rich in neutrons, like $^{18}$C with 6 protons and 12 neutrons.  And so we went for it by the very exotic pion double charge exchange (DCX) reaction $(\pi^+,\pi^-)$. We discovered $^{18}$C by the reaction $^{18}\mathrm{O}(\pi^-,\pi^+)^{18}\mathrm{C}$~\cite{nann}.  That was exciting. As I said before, exotica is addictive.  So, after $^{18}$C we went for $^9$He, 2~protons+7~neutrons, $(N-Z)/A= 5/9$, by means of the reaction $^9\mathrm{Be}(\pi^-,\pi^+)^9\mathrm{He}$~\cite{seth1}.  We found it, and Bethe called it \textbf{``a drop of neutron star''}.  How much more exotic can you get? Well, how about $^6$H by $^6\mathrm{Li}(\pi^-,\pi^+)^6\mathrm{H}$.  We tried and failed to find it, bound or unbound~\cite{parker}.

\textbf{So, running after exotica can lead to disappointments.}

The other end of exotic nuclei is the \textbf{superheavy nuclei}.  I have never worked in this field.  But Berkeley, Dubna, and GSI have crossed swords in claims about who has the heaviest of the superheavy.  After some embarrassing incidents, the current winner is $^{294}\mathrm{X}_{114}$ with 114 protons and 180~neutrons~\cite{ogan}.  That is exotic!

\section{Chasing Exotica in Quark Physics}

Quarks carry color, and \textbf{only color-neutral hadrons, $\bm{q\bar{q}}$ mesons or $\bm{qqq}$, baryons exist} in nature.  In the quark bag model~\cite{chodos} hadrons with other color-neutral combinations, such as $(qqq)(qqq)$ \textbf{dibaryons}, or $qq\bar{q}\bar{q}$ \textbf{four-quark} state can exist.  de Swart and colleagues calculated the masses of scores of dibaryons~\cite{mulders} and started a stampede for the search of dibaryons.

Lots of people started looking for dibaryons in their old experiments, analyzing old bubble chamber pictures and claiming observation of scores of dibaryons.  As many as 40 dibaryon states were claimed in the mass range $1900-2300$~MeV~(Fig.~1). We thought we could become famous by pinning these dibaryons down since we had orders of magnitude greater luminosity and energy resolution available at the Los Alamos Meson Factory.  Instead of becoming \textbf{famous} for discovering dibaryons, we became \textbf{infamous} for killing all of them. No Dibaryons anywhere in Fig.~2.

\begin{figure}[!tb]
\begin{center}
\includegraphics[width=2.1in]{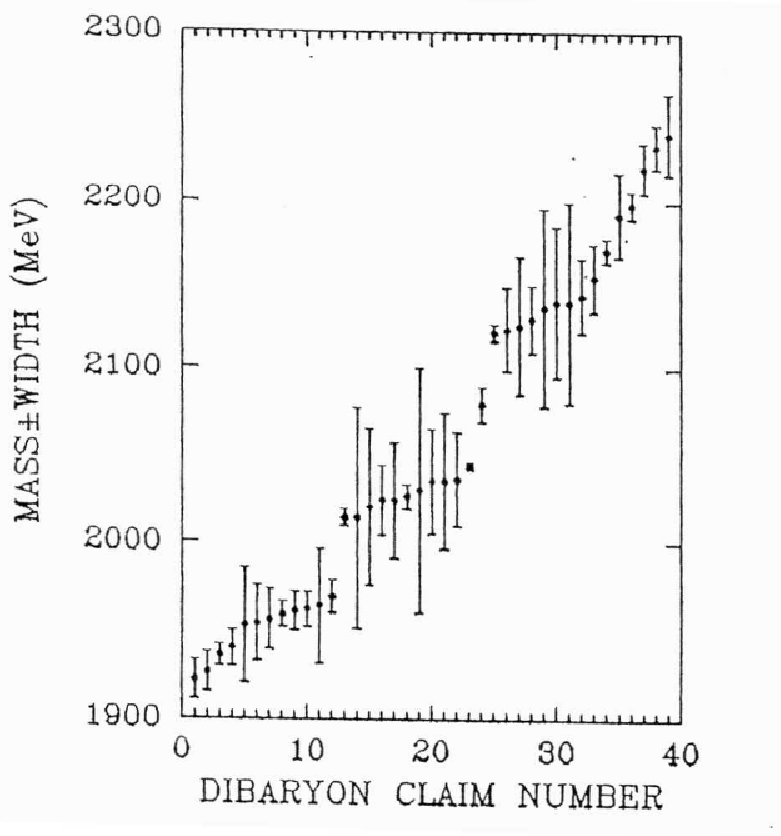}
\includegraphics[width=2.95in]{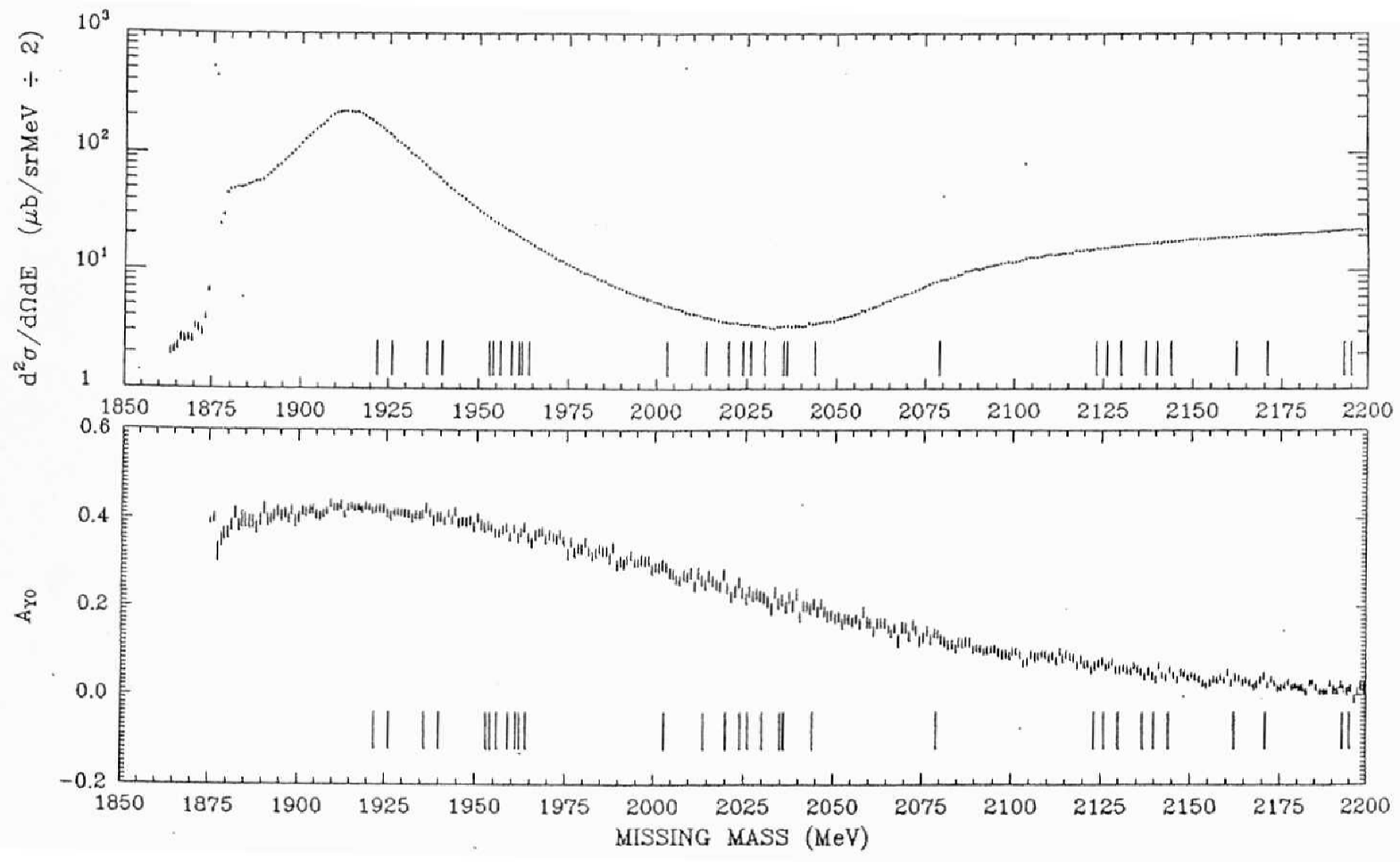}
\end{center}

\caption{ (Left) Claims for dibaryons.  The ordinate shows calimed mass (points) and widths (error bars).  (Right) $\vec{p}+d\to p'+X$ at $T_p=800$~MeV, $\Theta=15^\circ$: Vertical lines mark the masses of the dibaryons claimed.  (Upper panel) Differential cross sections.  (Lower panel) Analyzing powers.}
\end{figure}

\section{Pentaquark}

But that is not the end of this story.  If not two baryons making a dibaryon, how about a baryon$+$a meson, or a color-neutral pentaquark?  It surfaced a few years ago by the claim by Nakano et al. of a narrow peak, called $\Theta^+$, with a mass of $M(\Theta^+)=1540\pm10$~MeV, $\Gamma(\Theta^+)<25$~MeV, in the invariant mass of $K^+n$ in the reaction $\bm{\gamma n\to K^-(K^+n)}$~\cite{nakano}.  If true, it would have strangeness~+1 and at least five quarks/antiquarks.  The object was so exotic that a stampede of \textbf{confirming} claims flooded the literature.  An equal number of non-observations were reported.  If you go to Google, you find 99,800 entries for pentaquark (before this symposium), and it will be difficult to decide whether the pentaquark is alive or not.  

In a high-statistics repeat of their own measurement, JLab found that their own earlier observation of $\Theta^+$ was false and no evidence for the existence of the pentaquark exists~\cite{mckinnon}.  However, rumor has it that Nakano et al. claim that they still see the pentaquark in a high-statistics remeasurement. 

\textbf{So, once claimed, an exotic is difficult to kill!}  I end with a quote from PDG08 summarizing the saga of the pentaquark: \textit{``The whole story --- the discoveries themselves, the tidal wave of papers by theorists and phenomenologists that followed, and the eventual \textbf{`undiscovery'} --- is a curious episode in the history of science.''}

\section{Glueballs and Hybrids}

Since glue carries color, it is possible to have hadrons build of pure glue, called glueballs $\left|gg\right>$, and hybrid mesons containing glue, $\left|q\bar{q}g\right>$.  These have been predicted since the inception of QCD \cite{gellmann}.

Glueballs have generally the same $J^{PC}$ as $q\bar{q}$ mesons, and they mix with them.  It is therefore essentially impossible to find a pure glueball.  Nevertheless, brave searches and claims and counter--claims have been made.  The summary of the situation is that pieces of the $J^{PC}=0^{++}$ glueball are mixed into at least three well--known isoscalar mesons, $f_0(1370,1500,1710)$ and the pure exotic, $\left|gg,0^{++}\right>$ has been lost.  A tensor $J^{PC}=2^{++}$ glueball has had equally disappointing fate.  It has surfaced many times, but I believe it was firmly put to rest by us in a $p\bar{p}$ measurement at LEAR (CERN) \cite{amsler}.

Hybrids $\left|q\bar{q}g\right>$ have an advantage over glueballs.  They can have $J^{PC}=1^{-+},~2^{+-},\ldots$ which are not permitted for $q\bar{q}$ mesons.  Such objects are manifestly exotic.  In our $\pi^-p$ experiment (E852) at BNL we claimed to have discovered at least three $1^{-+}$ mesons $\pi_1(1400,1600,2000)$, and a $2^{+-}$ meson $h_2(1900)$~\cite{adamsexp}.  I have to admit that while these hadrons are definitely not $q\bar{q}$ mesons, they also admit the possibility of being four--quark states, and not hybrids.  In either case they are exotic.

\section{The H Dibaryon}

The $uuddss$ H dibaryon was predicted by Jaffe~\cite{jaffe}, but it became so exotic that it was even considered a candidate for dark matter.  Stubborn searches for the H were made for years at Brookhaven and KEK.
The $u,d$ quark dibaryons died a long time ago, but the H dibaryon lived longer.  By now, however, by common consensus it is also considered dead.  For a detailed history see~\cite{tralfner}.

\section{Exotica in QCD}

In Dec. 1974, a large narrow peak was disocvered at $\sim3.1$~GeV mass at Brookhaven and SLAC~\cite{aubert} in $e^+e^-$ formation and $\mu^+\mu^-$ decay.  It was the $\bm{J/\psi}$ which launched the era of modern Quantum Chromodynamics (QCD).
It is amusing to note that barely four weeks later \textbf{eight papers} by theoretical physicists (including four Nobel laureates) appeared in the Jan. 6, 1974 issue of Physical Review Letters~\cite{various}, offering explanations of what $J/\psi$ might be.  Several of them were truely exotic explanations, like $J/\psi$ was a bound state of a baryon/antibaryon, or two spin--one mesons, or it was a member of a $15\oplus1$ dimensional representation of SU(4).
\textbf{Tells you that nobody is immune to the seduction of exotica.}

I have been talking too much about the exotics which failed to materialize.  Let me now, for awhile, focus on exciting physics which is \textbf{not exotica, but excitica} (my construct for something very exciting).

\begin{figure}[!tb]
\begin{center}
\includegraphics[width=5.2in]{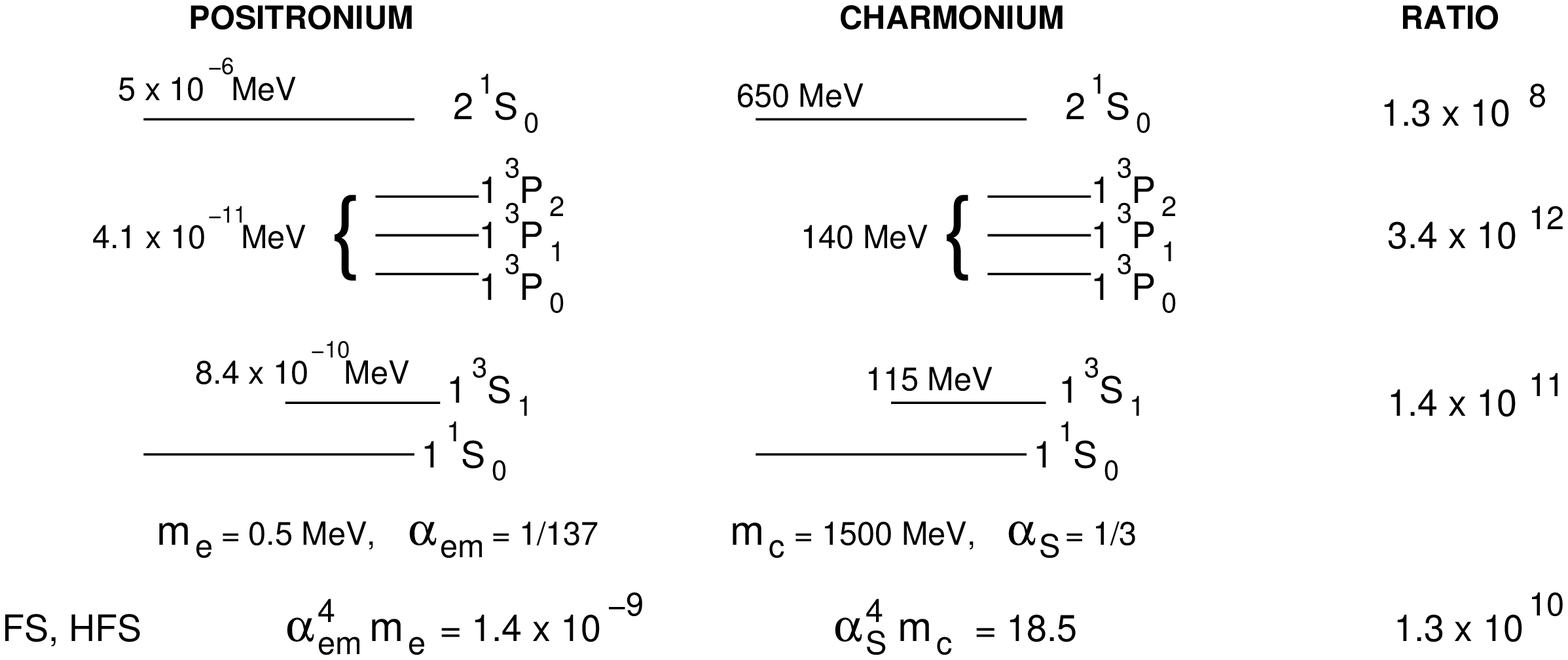}
\end{center}
\caption{Comparing Positronium ($e^+e^-$) and Charmonium ($c\bar{c}$) spectra.}
\end{figure}

\section{QCD versus QED}

The QCD potential which arises due to the exchange of a massless vector \textbf{photon} is $\bm{V(r)\propto\alpha_{em}/r}$.  The QCD potential due to the exchange of a massless vector \textbf{gluon} is $\bm{V(r)\propto\alpha_{\mathrm{strong}}/r}$.  Because free quarks do not exist, in QCD there is an additional confinement term proportional to  $\bm{r}$.

With such close analogy to QED, it is interesting to compare the QCD spectrum of charmonium with the QCD spectrum of positronium, with masses and interactions miles apart.  The similarity is nothing short of fantastic. \textbf{Nature repeats herself!} with energy scales different by a factor $\bm{\sim10^{10}}$.\\

\section{Hyperfine Interaction in QCD}

The Coulombic $(\propto1/r)$ part of the QCD interaction gives rise to the usual spin dependence in the potential, with spin-orbit, tensor, and spin-spin components, in addition to the central part.  Of these three, arguably the most important is the spin-spin interaction.  For example, the ground state masses of $q\bar{q}$ mesons are:
$$M(q_1\bar{q}_2) = m_1(q_1) + m_2(q_2) + A_{hf} \left[ \frac{\vec{s}_1\cdot\vec{s}_2}{m_1m_2} \right]$$
In order to determine the hyperfine interaction, $A_{hf}$, it is necessary to measure the \textbf{hyperfine splitting} between the spin--singlet and spin--triplet states.  This means identifying and measuring the massses of $^3L_J$ and $^1L_J$ states. The masses of \textbf{spin--triplet} $^3L_J$ states, $^3S_1$ and $^3P_J$ states are well-determined because either they are directly populated in $e^+e^-$ annihilation $(\left|^3S_1\right>)$ or they are reached by strong E1 transitions from the $\left|^3S_1\right>$ states $(\left|^3S_1\right>\to\gamma_{E1}\left|^3P_J\right>)$.  The \textbf{spin--singlet} states $^1L_{J=L}$ can not be directly formed, and radiative transitions to them from spin--triplet states are either forbidden or weak M1. The net result is that our knowledge of the spin--singlet states, and therefore of the hyperfine interaction, has been very poor in the past.  Very recently this has changed.

\begin{figure}[!tb]
\begin{center}
\includegraphics[width=2.4in]{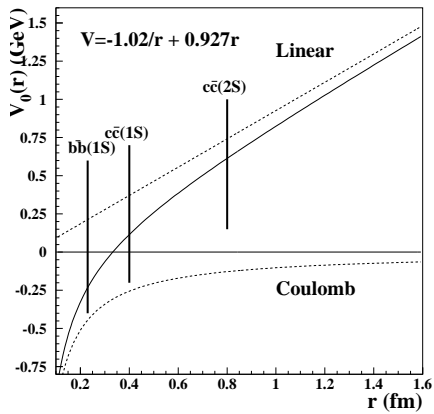}
\end{center}
\caption{The $S-$wave bound states of charmonium and bottomonium as a function of the radius of the $q\bar{q}$ potential.}
\vspace{-1.cm}
\end{figure}

\begin{figure}[!tb]
\begin{center}
\includegraphics[width=3.0in]{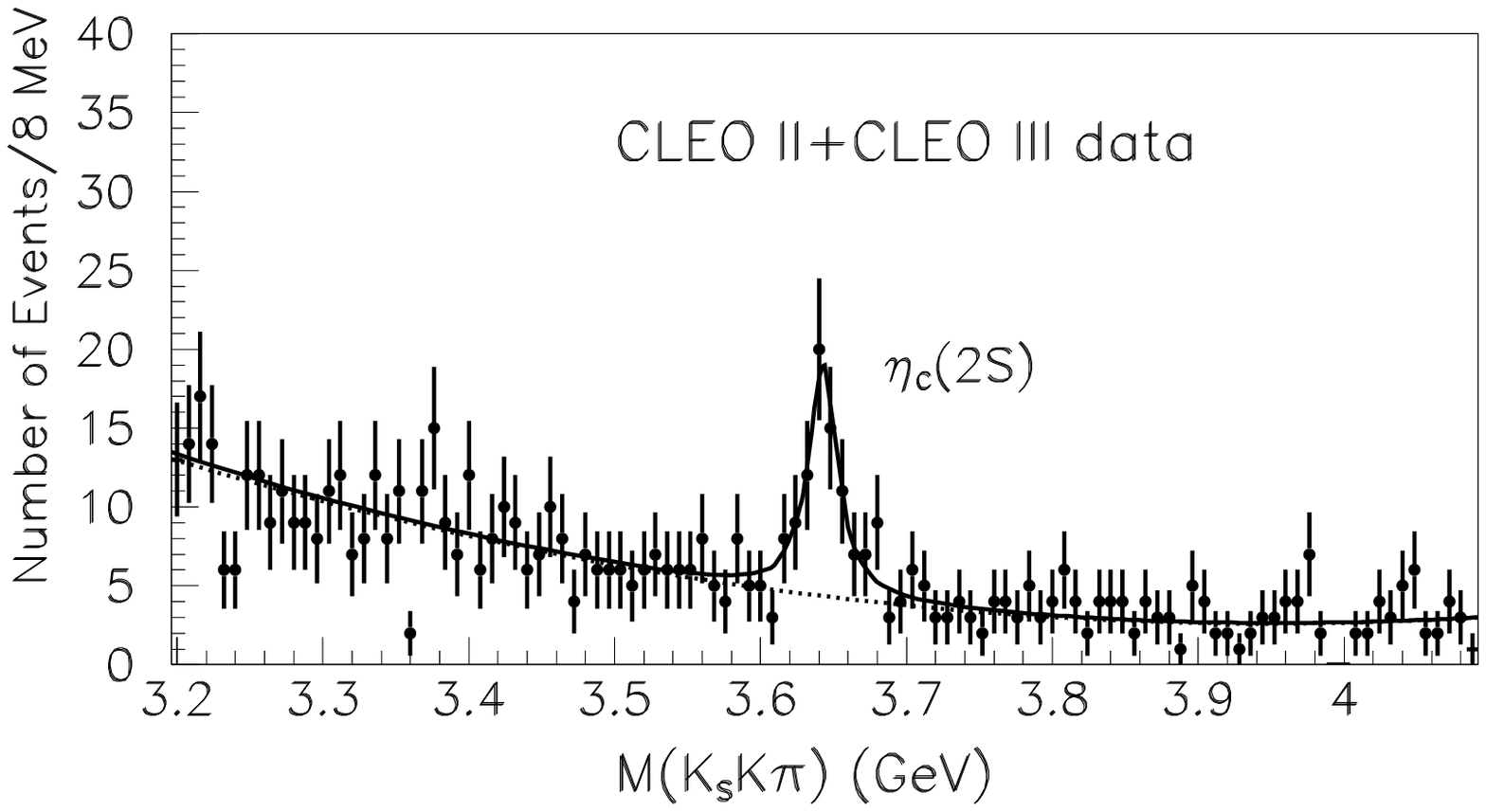}
\includegraphics[width=2.2in]{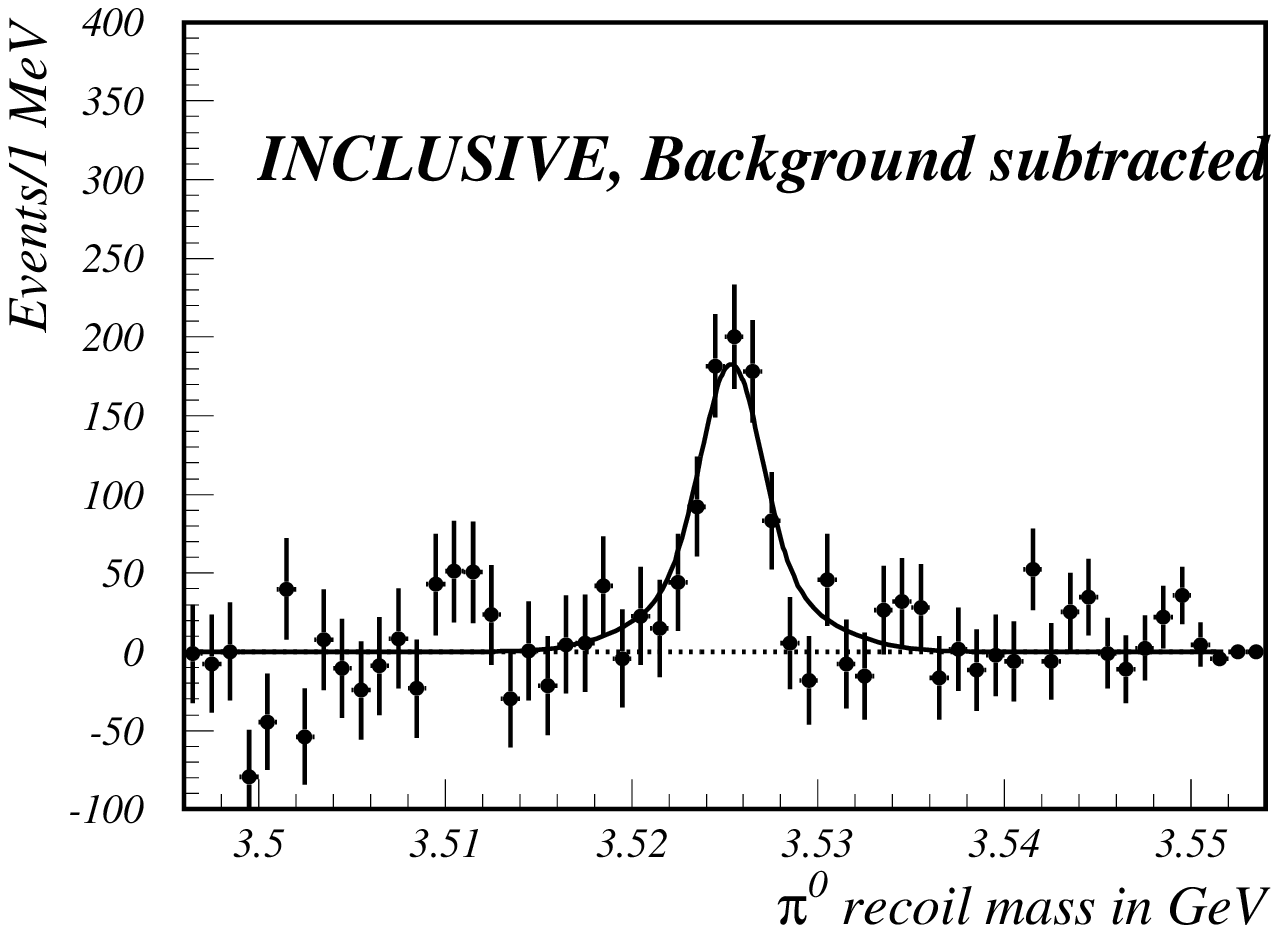}
\end{center}
\caption{Identification of charmonium spin--singlet states: (left) $\eta_c'(2^1S_0)$ produced in $\gamma\gamma$ fusion, (right) $h_c(1^1P_1)$ produced in $\psi'\to\pi^0h_c$. }
\end{figure}

For heavy quark systems, $c\bar{c}$ charmonium, and $b\bar{b}$~bottomonium, we would like to know how the hyperfine interaction changes as we move from the Coulomb dominated region of the $q\bar{q}$~potential to the confinement dominated region.  We would like to study the change in the hyperfine interaction
\begin{enumerate}
\item between $c\bar{c}(1S)$ and $c\bar{c}(2S)$
\item between $c\bar{c}(1S)$ and $b\bar{b}(1S)$
\item between $c\bar{c}(1S)$ and $c\bar{c}(1P)$
\end{enumerate}


Until recently, the only hyperfine splitting known was for the charmonium $1S$ states (see Fig.~3)
$$\Delta M_{hf}(1S)\equiv M(J/\psi(1S)) - M(\eta_c(1S)) = 116.7\pm1.2~\mathrm{MeV}$$
\textbf{$\bm{\eta_c'(1^1S_0)}$, $\bm{h_c(1^1P_1)}$, and $\bm{\eta_b(1^1S_0)}$} were not even identified.  In the last five years, all this has changed due to the measurements at Belle, BaBar, and CLEO.  The results are:
\begin{align*}
\Delta M_{hf}(2S)_{c\bar{c}} & \equiv M(\psi'(2S))-M(\eta_c'(2S)) = +43.2\pm3.4~\mathrm{MeV}~(\mathrm{CLEO}~\cite{asner})\\
\Delta M_{hf}(1P)_{c\bar{c}} & \equiv M(\left<\chi_{cJ}(1P)\right>)-M(h_c(1P)) = +0.02\pm0.23~\mathrm{MeV}~(\mathrm{CLEO}~\cite{rosner})
\end{align*}
Even more recently BaBar has claimed the identification $\eta_b(^1S_0)$ with the result
$$\Delta M_{hf}(1S)_{b\bar{b}} \equiv M(\Upsilon(1S))-M(\eta_b(1S)) = +71.4^{+4.1}_{-3.5}~\mathrm{MeV}~(\mathrm{BaBar}~\cite{aubert2})$$
An overall understanding of these hyperfine splittings is going to be a challenge to the theorists.

\section{CHARMONIUM EXOTICS: The Unexpected States Above the $D\overline{D}$ Threshold}

I now return to the domain of Exotica. Recently, a number of new states have been claimed in the mass region 3800--4700~MeV, above the $D\overline{D}$ breakup of charmonium at 3730~MeV.
Three years ago, all that was known above $D\overline{D}$ was four vector states $\psi(3770,~4040,~4160,~\mathrm{and}~4415)$ observed as enhancements in the ratio, $R = \sigma(hh)/\sigma(\mu^+\mu^-)$. However, the great excitement, often called the \textbf{renaissance} in hadron spectroscopy, has come from the recent discovery of a whole host of unexpected states by the meson factory detectors, Belle and BaBar.  

The new states are called \textbf{``charmonium-like states''}, not because they naturally fit into the spectrum of charmonium states, but because they seem to always decay into final states containing a charm quark and an anti-charm quark.  There are at least eleven of them around.  The alphabet soup is getting thick with reports of X(3872), Y(4260), Y(4361), Y(4660), X(3940), Y(3940), Z(3940), X(4160), Z$^\pm$(4430), Z$^\pm_1$(4051) and Z$^\pm_2$(4248).  Except for the first two, X(3872) and Y(4260), which have been observed in measurements at several laboratories, the remaining nine come exclusively from Belle.  They have not been reported by BaBar with similar capabilities, and in two cases, Y(4325) and the $Z^\pm$, they have been contradiected by BaBar.  Reminds you of the old dibaryon story.  I do not want to express my skepticism any further, but tell you only about the two certain exotics, X(3872) and Y(4260).

\subsection{X(3872) and the molecular model}

This narrow state with {$M(\mathrm{X}) = \bm{ 3872.2 \pm 0.8}$~MeV}, and $\Gamma(\mathrm{X})=1.34\pm0.64$~MeV,  has been observed by Belle, BaBar, CDF, D\O, and it definitely exists.  [PDG08]
CDF angular correlation studies  show that its $J^{PC} = 1^{++}$ or $2^{-+}$.
X(3872) does not easily fit in the charmonium spectrum. 
Since its mass is very close to $M(D)+M(D^*)$, the most popular conjecture is that it is a weakly \textbf{bound molecule} of $D$ and $D^*$. If so, our recent precision measurement of $D^0$ mass at CLEO gives 
{$M(D^0D^{0*})=\bm{3871.81\pm0.36}$~MeV}.  This corresponds to X(3872) being \textbf{ unbound} by $0.4\pm0.8$~MeV.  If X(3872) were even bound by $\sim0.4$~MeV,
 the branching fraction for the molecule's breakup into $D\overline{D}\pi$ is predicted to be factor 400 smaller than observed.  These observations raise serious doubts about the molecular model for X(3872).

Stop the presses: CDF now reports~\cite{cdf} $M(\mathrm{X})=3871.46\pm0.19$~MeV.  So we now have X(3872) bound by $0.35\pm0.41$~MeV. The problem of the almost-bound/unbound nature of X(3872) is getting more and more sharply defined, and it is getting to be more and more exotic.

\subsection{Y(4260) and the strange Vector}

The Y(4260) has been observed in ISR production by BaBar, CLEO and Belle, and in direct production by CLEO.  Y(4260) is clearly \textbf{ a vector} with $J^{PC}=1^{--}$.  All known charmonium vectors are seen prominently as huge enhancements i hadronic decays, usually measured as the ratio $R=\sigma(h^+h^-)/\sigma(\mu^+\mu^-)$. But this vector is a very strange one, since it sits at a very deep minimum in R, with
$$M(\mathrm{Y}(4260))=4263^{+8}_{-9}~\mathrm{MeV},\quad\Gamma(\mathrm{Y}(4260))=95\pm14~\mathrm{MeV}\qquad(\mathrm{PDG}08)$$
So it is not likely to be a charmonium vector, which are all spoken for, anyway. 
So what is Y(4260)?

It is suggested that Y(4260) is a $c\bar{c}g$ charmonium hybrid. If so, there ought to be $0^{-+}$ and $1^{-+}$ hybrids companions nearby. Where are they? It is a real experimental challenge to clarify this situation before taking any theoretical conjecture seriously.

\section{Epilogue}

The sum total of the experiences in this journey through hadronic \textbf{exotica} is that the journey is certainly worth it.  It is unquestionably exciting.  But the road is full of pitfalls and disappointments.
\begin{center}
\textbf{Only the brave should enter!}
\end{center}
They should be proud of their successes, and humble enough to admit their failures.

\end{document}